\documentclass[preprint,pra,aps,showpacs,preprintnumbers,amsmath,amssymb,eqsecnum]{revtex4} 

\usepackage{graphicx}
\usepackage{dcolumn}
\usepackage{bm}
\usepackage{verbatim}
\usepackage{epsfig}
\usepackage{epstopdf}

\def\ket#1{\! \! \mid#1\rangle}
\def\mel#1#2#3{\langle{#1}\!\mid\!{#2}\!\mid \!{#3}\rangle}  
\def\Re{{\rm Re}}

\begin{document}


\title{A quasi-probability for the arrival time problem with links to 
backflow and the Leggett-Garg inequalities}

\author{J.J.Halliwell}%

\email{j.halliwell@imperial.ac.uk}

\author{H.Beck}

\author{B.K.B.Lee}

\author{S.O'Brien}

\affiliation{Blackett Laboratory \\ Imperial College \\ London SW7
2BZ \\ UK }



\begin{abstract}
The arrival time problem for the free particle in one dimension may be formulated as the problem of determining a joint probability for the particle being found on opposite sides of the $x$-axis at two different times.
We explore this problem using a two-time quasi-probability linear in the projection operators, a natural counterpart of the corresponding classical problem. We show that it can be measured either indirectly, by measuring its moments in different experiments, or directly, in a single experiment using a pair of sequential measurements in which the first measurement is weak (or more generally, ambiguous). We argue that when positive, it corresponds to a measurement-independent arrival time probability. For small time intervals it coincides approximately with the time-averaged current, in agreement with semiclassical expectations.
The quasi-probability can be negative and we exhibit a number of situations in which this is the case. We interpret these situations as the presence of ``quantumness'', in which the arrival time probability is not properly defined in a measurement-independent manner.
Backflow states, in which the current flows in the direction opposite to the momentum, are shown to provide an interesting class of examples such situations.
We also show that the quasi-probability is closely linked to a set of two-time Leggett-Garg inequalities, which test for macroscopic realism.

\end{abstract}


\maketitle



\newcommand\beq{\begin{equation}}
\newcommand\eeq{\end{equation}}
\newcommand\bea{\begin{eqnarray}}
\newcommand\eea{\end{eqnarray}}

\def\A{{\cal A}}
\def\D{\Delta}
\def\H{{\cal H}}
\def\E{{\cal E}}
\def\p{\partial}
\def\la{\langle}
\def\ra{\rangle}
\def\ria{\rightarrow}
\def\x{{\bf x}}
\def\y{{\bf y}}
\def\k{{\bf k}}
\def\q{{\bf q}}
\def\p{{\bf p}}
\def\P{{\bf P}}
\def\r{{\bf r}}
\def\s{{\sigma}}
\def\a{\alpha}
\def\b{\beta}
\def\e{\epsilon}
\def\U{\Upsilon}
\def\G{\Gamma}
\def\om{{\omega}}
\def\Tr{{\rm Tr}}
\def\ih{{ \frac {i} { \hbar} }}
\def\trho{{\rho}}
\def\au{{\underline \alpha}}
\def\bu{{\underline \beta}}
\def\pp{{\prime\prime}}
\def\id{{1 \!\! 1 }}
\def\half{\frac {1} {2}}
\def\jjh{j.halliwell@ic.ac.uk}

\section{Introduction}

The arrival time problem in quantum mechanics has been the subject of many papers over the years
\cite{Allcock,TQM,TQM2,rev1,rev2}. It continues to be an interesting problem since it is a simple example of a question easily addressed in classical mechanics whilst quantum mechanics provides no unique answer. Most simply stated, in one-dimensional quantum mechanics, it is the question of determing the probability that an incoming wave packet crosses the origin during a given time interval.

There are many approaches to this problem. One of the earliest and most-studied appraoches involves the construction of an arrival time operator \cite{AB} by quantizing the classical result $-mx/p$ (for an incoming classical particle with momentum $p$ and position $x$). This typically leads to arrival time operators which are not self-adjoint \cite{Pauli}, although self-adjoint variants have been proposed \cite{DM,MI,HELM}. Even when self-adjoint,  such operators are not obviously connected to a particular measurement scheme, since there is no obvious way of creating a physically realizable coupling between a measuring device and any of the proposed arrival time operators. (See however, Ref.\cite{HELM}).

In the present paper we will focus on arrival times in one-dimensional quantum mechanics defined using sequential measurements onto the positive or negative $x$-axis. Suppose we consider a system prepared in state $|\Psi \rangle$ at $t=0$ and use measurements at a set of closely spaced times $t_1, t_2, \cdots t_n$ to determine its behaviour. Introducing projection operators $P_+ = \theta ( \hat x)$ and $P_- = \theta (-\hat x) $ onto the positive and negative $x$-axis, the state
\beq
|\Psi_{123 \cdots n} \rangle = P_+ (t_n) P_- (t_{n-1}) \cdots P_-(t_2) P_- (t_1) \ | \Psi \rangle,
\label{1.1}
\eeq
represents the amplitude for the history in which the particle lies  in $x<0$ at times $t_1, t_2, \cdots t_{n-1} $ and is in $x>0$ at $t_n$, where $P_+(t)$ denotes the projection operator in the Heisenberg picture. For sufficiently close spacing of the times, this object is then a plausible candidate for the amplitude for the particle to make a left-right crossing of the origin, for the first time, during the time interval $[t_{n-1},t_n]$. The probability for the crossing is then the norm of this state.

Note that in an expression of the form Eq.(\ref{1.1}), one would expect the Zeno effect \cite{Zeno} to come into play for sufficiently frequent measurements.  This is indeed the case -- it becomes significant when the time interval between projectors is smaller than $\hbar / E$, where $E$ is the energy scale of the incoming packet \cite{HaYe1,HaYe5}.

The probability derived from Eq.(\ref{1.1}) is a special case of the standard quantum-mechanical formula for a set of $n$ sequential measurements at $n$ times,
\beq
p(s_1, s_2, \cdots s_n) = {\rm Tr} \left( P_{s_n} (t_n) P_{s_{n-1}} (t_{n-1}) \cdots  P_{s_1} (t_1) \rho P_{s_1} (t_1) \cdots P_{s_{n-1}} (t_{n-1} ) \right),
\label{1.2}
\eeq
where we have introduced the convenient notation in which the projectors $P_+$ and $P_-$ are written
\beq
P_s = \half \left( 1 + s \hat Q \right),
\eeq
where $\hat Q = {\rm sign} ( \hat x) $ and $s = \pm 1$ (and the introduction of the dichomotic variable $Q$ makes notational contact with the related papers Refs.\cite{HalLG1,HalLG2,HalLG3,HalLG4}). We use the Heisenberg picture in which $P_s (t)= e^{\ih Ht} P_s e^{- \ih Ht} $, ${\rm Tr}$ denotes the trace, and we have taken a general mixed initial state $\rho$.

The formula Eq.(\ref{1.2}) gives the probabilities for the set of all possible histories in which the particle may be in $x<0$ or $x>0$ at each of the times $t_1, t_2 \cdots t_n $. This will permit a quite detailed characterization of the arrival time probability. 

In what follows we will focus on the simplest case in which measurements onto the positive and negative $x$-axis are made at just two times, $t_1$, $t_2$. The probabilities for the four possible histories of the system are then given by the two-time version of Eq.(\ref{1.2}), which we denote,
\beq
p_{12}(s_1, s_2) = {\rm Tr} \left( P_{s_2} (t_2) P_{s_1} (t_1) \rho P_{s_1} (t_1) \right).
\label{2timep}
\eeq
From this we may obtain the probabilities for, respectively, left-right and right-left crossings, $p(-,+)$ and $p(+,-)$, and for remaining on the left or on the right, $p(-,-)$ and $p(+,+)$.

In general, sequential measurements in quantum mechanics have the property that each measurement disturbs any later measurement. To quantify this, consider the probability for a measurement at the second time only in which no earlier measurement was carried out, namely
\beq
p_2 (s_2) = {\rm Tr} \left( P_{s_2} (t_2) \rho \right).
\label{single}
\eeq
If the first measurement in Eq.(\ref{2timep}) does not disturb the later measurement, then one would expect that a relation of the form
\beq
\sum_{s_1} p_{12} (s_1, s_2) = p_2 (s_2),
\label{cond}
\eeq
would hold. This is often called a probability sum rule (or, elsewhere, the ``no signaling in time" condition \cite{KoBr,Cle}). However it does not hold in general since we have
\beq
\sum_{s_1} p_{12} (s_1, s_2 ) = {\rm Tr} \left( P_{s_2} (t_2)  \rho_M (t_1) \right),
\eeq
where $\rho_M (t_1) $ denotes the measured density operator,
\beq
\rho_M (t_1) = \sum_{s_1} P_{s_1} (t_1) \rho P_{s_1} (t_1).
\label{rhoM}
\eeq
This does not coincide with Eq.(\ref{single}) except under very specific conditions, such as an initial density matrix which is diagonal in the appropriate basis.

This feature of the two-time sequential measurement probabilities means that it is difficult to define the arrival time probability in this way. Of course one could still physically measure Eq.(\ref{2timep}) and it is still a probability, in the sense that its components all sum to $1$. But the failure to satisfy the sum rule suggests that such a probability formula has in general significant dependence on the measurement procedure employed to determine it and indeed specific models exhibit exactly this property. (It is however, sometimes possible to extract the ideal arrival time distribution proposed by Kijowski \cite{Kij} from a particular measurement process \cite{DEHM}).

Furthermore, there is a related question around the issue of ``quantumness''. The sum rule Eq.(\ref{cond}) is, as we shall see, a stringent classicality condition since it effectively requires zero interference between different histories. This suggests that the two-time probability formula only gives sensible results in a highly classicalized regime. However, this is on the face of it unusually restrictive, constrasting strongly with the operator approach to the arrival time problem, where there does not appear to be any restrictions on the situations in which arrival time probabilities can be found.
Some reasonable questions to ask are therefore as follows: Is it still possible to assign probabilities to two-time histories in a physically sensible and measurement-independent way in the face of non-trivial interference? 
In what ways does quantumness show itself in the arrival time problem, other than the failure of the sum rule Eq.(\ref{cond})? How is it measured?

In this paper we will offer interesting possible answers to these questions.
The key point is that in attempting to find two-time probabilities for a pair of non-commuting observables (namely, the sign of $\hat x$ at two different times), quantum mechanics offers more than one possibility.
Different ways correspond to different measurement methods.
The first obvious way is via sequential measurements, as described. But there is a second, different sort of approach which is to employ the closely related two-time quasi-probability,
\beq
q(s_1, s_2) = \half {\rm Tr} \left( \left( P_{s_2} (t_2) P_{s_1} (t_1) +  P_{s_1} (t_1) P_{s_2} (t_2) \right) \rho  \right).
\label{quasi}
\eeq
Expressions of this type were proposed in Ref.\cite{GoPa} and explored in Refs.\cite{HalLG1,HaYe2}.
This expression has the advantage that it satisfies sum rules of the form Eq.(\ref{cond}), but can be negative, since the projectors at different times do not commute. However, it is clearly a natural analogue of the corresponding classical problem and furthermore, its possible negativity provides exactly the indicator of quantumness that we seek.

The purpose of this paper is to explore the properties of the quasi-probability Eq.(\ref{quasi}) as both an indicator of quantumness in the arrival time problem (and beyond) and, when positive, a candidate probability for it.

We start in Section 2 by discussing the general properties of a quasi-probability of the form Eq.(\ref{quasi}), for a general dichotomic variable $Q$ and its relationship to the two-time measurement formula Eq.(\ref{2timep}). We will show that it can be positive as long as quantum inteference effects are suitably bounded, a relaxation of the sum rule Eq.(\ref{cond}) (which requires zero interference). 
We also show that the quasi-probabilty Eq.(\ref{quasi}) cannot be smaller than $ - \frac{1}{8}$.

In Section 3 we will show that the quasi-probability may be measured experimentally either indirectly, by measuring its moments, or more directly using sequential measurements that are weak or, more generally, ambiguous measurements.
In Section 4 we show that the quasi-probability is closely related to the current and to the kinetic energy density in the short time limit, thereby linking to standard results for the arrival time distribution.

Sections 5, 6 and 7 explore the types of states which lead to negative quasi-probability. In Section 5
we explore the relationship between the quasi-probability and quantum backflow -- the unusual effect in which a state of positive momenta can have negative current. We find that the $q(-,+)$ component of the quasi-probability is negative for backflow eigenstates. These results also indicate a possible way of measuring quantum backflow.  In Section 6 we show how a significantly negative quasi-probability may be obtained using superpositions of gaussians and also that negativity can be obtained from a single gaussian.
In Section 7, we write the quasi-probability in terms of the Wigner-Weyl representation (of which the Wigner representation of density matrices is an example). This gives  a convenient phase space representation of the quasi-probability which makes it relatively easy to identify what is required of states which make the quasi-probability negative.

As stated above, the quasi-probability is regarded as a candidate probability for the arrival time problem, when non-negative. However, to be a genuine probability it must have a relationship to relative frequencies of certain outcomes in a measurement process. We explain this connection in Section 8.

In Section 9 we show that the condition $q(s_1,s_2) \ge 0 $ has the form of a set of two-time Leggett-Garg inequalities, which are tests of macrorealism, a very specific notion of classicality, analogous to local realism in Bell tests \cite{LG1,ELN}. We will outline how the measurements of the quasi-probabity described in earlier sections may be modified to meet the requirements of a Leggett-Garg test.
We summarize and conclude in Section 10.



\section{General Properties of the Quasi-probability.}

In this section we summarize some of the general properties of the quasi-probability Eq.(\ref{quasi}).
The properties described below concern a general dichomatic variable $Q$. Properties relating to the specific form of $Q$ relating to the arrival time problem, $Q ={\rm sign}(x)$, will be described later.

Because it is linear in both projection operators, the quasi-probability satisfies the relations,
\bea
\sum_{s_1} q(s_1, s_2) &=& {\rm Tr} \left( P_{s_2} (t_2) \rho \right) = p_2(s_2),
\label{con1}
\\
\sum_{s_2} q(s_1, s_2) &=& {\rm Tr} \left( P_{s_1} (t_1) \rho \right) = p_1(s_1).
\label{con2}
\eea
So unlike the two-time measurement probability Eq.(\ref{2timep}), it satisfies the probability sum rules.

Eq.(\ref{quasi}) is certainly not the only quasi-probability that matches the single time measurement marginals. One could for example add another term involving the commutator of the two projectors.
There is also some similarity with Wigner function constructions for finite dimensional systems \cite{Woo}. However, what distinguishes a particular choice of quasi-probability is the choice of measurement method and, as we shall see, the choice. Eq.(\ref{quasi}) has a natural link to weak measurements.

Furthermore, the quasi-probability Eq.(\ref{quasi})  has a simple relation to the standard quantum-mechanical two-time probability Eq.(\ref{2timep}), namely,
\beq
q(s_1, s_2) = p_{12} (s_1, s_2)  + {\rm  Re} D (s_1,s_2 |-s_1,s_2),
\label{qp}
\eeq
where the quantity
\beq
D (s_1,s_2 |s_1',s_2) = {\rm Tr} \left(  P_{s_2} (t_2) P_{s_1} (t_1) \rho P_{s_1'} (t_1)  \right),
\label{DF}
\eeq
is the so-called decoherence functional. Its off-diagonal terms are measures of interference between the two different quantum histories represented by sequential pairs of projectors.
(We use here the mathematical language of the decoherent histories approach \cite{GH1,GH2,GH3,Gri,Omn1,Hal2,Hal3,DoK,Ish,IshLin} but this is not a decoherent histories analysis of the arrival time problem. Such an analysis was carried out in Ref.\cite{HaYe5}.)
When
\beq
{\rm  Re} D (s_1,s_2 |s_1',s_2) = 0, \ \ \ \ {\rm for} \ \ s_1 \ne s_1',
\eeq
a condition normally referred to as consistency,
there is no interference and we have $q(s_1,s_2) = p(s_1,s_2) $, and the sum rule Eq.(\ref{cond})
is satisfied exactly. However, noting that $p(s_1,s_2) $ is always non-negative, we see from Eq.(\ref{qp}) that $q(s_1, s_2)$ will be non-negative if the off-diagonal terms of the decoherence functional are bounded,
\beq
 \left| {\rm  Re} D (s_1,s_2 |-s_1,s_2) \right| \le p_{12} (s_1, s_2).
\eeq

The requirement that the quasi-probability Eq.(\ref{quasi}) is non-negative
\beq
q(s_1, s_2) \ge 0,
\eeq
was named ``linear positivity" by Goldstein and Page and is one of the weakest conditions under which probabilities can be assigned to non-commuting variables, subject to agreeing with the expected formulae for commuting projectors and to matching the probabilities for projectors at a single time \cite{GoPa}.
(See also Ref.\cite{HaYe2} for other weak probability assignment conditions).
It is satisfied very easily in numerous models, for suitably chosen ranges of parameters, since it requires only partial suppression of quantum interference, not complete destruction of it.

The quasi-probability is very conveniently expanded in terms of its moments,
\beq
q(s_1,s_2) = \frac {1}{4} \left(1 + \langle \hat Q_1 \rangle  s_1 +  \langle \hat Q_2 \rangle s_2  + C_{12} s_1 s_2 \right),
\label{mom}
\eeq
where the correlation function $C_{12}$ is given by,
\beq
C_{12} = \half \langle \hat Q_1 \hat Q_2 + \hat Q_2 \hat Q_1 \rangle.
\label{corr}
\eeq
(See Refs.\cite{HaYe2,Kly} for more on this useful representation). Here, we use the convenient notation
$\hat Q_1 = \hat Q(t_1)$ and $\hat Q_2 = \hat Q(t_2)$.
By contrast the two-time measurement probability, which is always non-negative, has the form
\beq
p_{12} (s_1,s_2) = \frac {1}{4} \left(1 + \langle \hat Q_1 \rangle  s_1 +  \langle \hat Q_2^{(1)} \rangle s_2  + 
C_{12} s_1 s_2
\right).
\label{momp}
\eeq
Here, $\langle \hat Q_2^{(1)} \rangle$ denotes the average of $Q$ at time $t_2$ in the context in which an earlier meausrement was made at $t_1$ and is given explicitly by
\beq
\langle \hat Q_2^{(1)} \rangle = \langle \hat Q_2 \rangle + \half \langle [\hat Q_1, \hat Q_2] \hat Q_1 \rangle
\label{extra}
\eeq
This extra term on the right-hand side, which clearly vanishes when $\hat Q_1$ and $\hat Q_2$ commute,
 is in fact the only difference between $q(s_1,s_2)$ and $p(s_1,s_2)$ and in particular note that the quasi-probability and the two-time measurement probability have the same correlation function,
\beq
C_{12} = \sum_{s_1, s_2} s_1 s_2 p_{12} (s_1, s_2 ) = \sum_{s_1, s_2} s_1 s_2 q(s_1, s_2 ).
\label{sameC}
\eeq
as previously noted \cite{Fri}.

Finally, one might reasonably ask how negative the quasi-probability may become in quantum theory. This is readily proved using the simply identity,
\beq
q(s_1, s_2) = \frac{1}{8}   \langle (1 + s_1 \hat Q_1 + s_2 \hat Q_2)^2 - 1 \rangle,
\label{idn}
\eeq
from which it is easily seen that
\beq
q(s_1, s_2) \ge - \frac{1}{8}.
\label{LB}
\eeq

\section{Measurement of the Quasi-probability}

There are two obvious methods for determining the quasi-probability experimentally.

\subsection{Reconstruction from Measurement of the Moments}

The first is to simply determine the moments of $q(s_1,s_2)$ from a number of different experiments and then assemble the quasi-probability via the moment expansion Eq.(\ref{mom}). This could be done for example, by making use of the sequential measurement probability, which, via the moment expansion Eq.(\ref{momp}) yields 
$\langle \hat Q_1 \rangle $ and $C_{12}$, and then $\langle \hat Q_2 \rangle $ could be determined in a second experiment measuring $Q$ at time $t_2$ only.

\subsection{Ambiguous Measurements}

A more direct alternative method is to use ambiguous measurements \cite{Amb,Ema}, which include weak measurements \cite{weak0} as a special case. We follow the convenient formulation given by Emary \cite{Ema}.

An ambiguous measurement relates to situations in which there is some sort of intrinsic noise in the system so that instead of determining a definite value $s$, the measurement leads to a value $\alpha$, with conditional probability $c_{\alpha s}$ (which is assumed to be known). Unlike conventional measurements, a repeated measurement may yield a different answer each time, constrained only by the conditional probability. In general $\alpha$ may run over a range greater than that of $s$, but here we assume that both take just two values. For a single time measurement, the desired probability of $s$, which we denote $\tilde p(s)$, may be inferred indirectly from the ambiguous measurement probability $p(\alpha)$, using
\beq
\tilde p(s) = \sum_{\alpha} d_{s \alpha} p(\alpha),
\eeq
where we have introduced the inverse $d_{s \alpha}$, which satisfies
\beq
\sum_{\alpha} d_{ s \alpha} c_{\alpha s'} = \delta_{ s s'}.
\eeq
(It is the left inverse in the more general case where $\alpha$ and $s$ run over different ranges).
Ambiguous measurements in quantum mechanics are described by POVM's of the form
\beq
F_{\alpha} = \sum_s c_{\alpha s} P_s,
\eeq
for which the probability is $ p(\alpha) = {\rm Tr} ( F_{\alpha} \rho) $.
The inferred probability for $s$ is then readily seen to be $\tilde p(s) = {\rm Tr} ( P_s \rho)$, the expected result.

Ambiguous measurements yield more subtle results when used as part of a sequence. We consider a situation in which there are two measurements at times $t_1$, $t_2$, in which the first measurement is ambiguous. We first introduce the operator $M_{\alpha}$, where $M^2_{\alpha} = F_{\alpha}$, so
\beq
M_{\alpha} = \sum_s \sqrt{ c_{\alpha s}} P_s.
\eeq
The two time joint probability of an ambiguous measurement obtaining a value $\alpha$ followed by a projective measurement obtaining a value $s_2$ is
\beq
p(\alpha,s_2)={\rm Tr} \left(P_{s_2}(t) M_\alpha (t_1)  \rho M_\alpha (t_1) \right).
\label{ambp}
\eeq
The inferred joint probability is then
\bea
\label{capq}
\tilde p (s_1,s_2) &=&\sum_\alpha d_{s_1\alpha}\ p(\alpha,s_2)
\nonumber \\
&=&\sum_\alpha  \sum_{s,s'} d_{s_1\alpha} \sqrt{c_{\alpha s}} \sqrt{c_{\alpha s'}}\, \text{Tr}\left( P_{s_2}(t_2) P_s(t_1)  \rho P_{s'} (t_1) \right).
\eea
Noting that the $s=s'$ terms in the trace on the right-and side yield the probability $p_{12} (s,s_2)$, and the $s \ne s'$ terms yield the decoherence functional Eq.(\ref{DF}), we have
\beq
\tilde p (s_1,s_2) = p_{12} (s_1, s_2) +
\sum_{s \ne s'}  \left( \sum_\alpha d_{s_1\alpha} \sqrt{c_{\alpha s}} \sqrt{c_{\alpha s'}} \right)
\ D (s, s_2 | s', s_2 ).
\label{pD}
\eeq

To proceed further we need to make a specific choice of POVMs. The following pair of POVMs are convenient for our purposes:
\bea
{F}_+&=&\frac{(1+\varepsilon)}{2}P_++\frac{(1-\varepsilon)}{2}P_- ,
\\
{F}_-&=&\frac{(1-\varepsilon)}{2}P_++\frac{(1+\varepsilon)}{2}P_-,
\eea
where $ 0 \le \varepsilon \le 1$.
These clearly correspond to projective measurements as $\varepsilon \rightarrow 1$ and to weak measurements as $\varepsilon \rightarrow 0 $. The matrices corresponding to the coefficients $c_{\alpha s}$ and $d_{s \alpha}$ are then,
\begin{align}
\textbf{c}=
\frac{1}{2} \begin{pmatrix}
1+\varepsilon & 1-\varepsilon\\
1-\varepsilon & 1+\varepsilon
\end{pmatrix}, && \textbf{d}=
\frac{1}{2 \varepsilon}\begin{pmatrix}
1+\varepsilon & -1+\varepsilon\\
-1+\varepsilon & 1+\varepsilon
\end{pmatrix}.
\label{mat}
\end{align}

The coefficient in Eq.(\ref{pD}) is now readily evaluated with the result,
\beq
\tilde p_\varepsilon (s_1,s_2) = p_{12} (s_1, s_2) +\sqrt{1-\varepsilon^2} \ {\rm Re} D(s_1,s_2|-s_1,s_2).
\eeq
Using the relation Eq.(\ref{qp}), this is now readily rewritten in the convenient form,
\beq
\tilde p_\varepsilon(s_1, s_2) =  (1-\sqrt{1-\varepsilon^2}) \ p_{12} (s_1, s_2) +\sqrt{1-\varepsilon^2}\ q(s_1,s_2).
\label{main}
\eeq
This simple and appealing result shows how the inferred probability from ambiguous measurements mediates continuously from the usual projective measurement result $p_{12}(s_1,s_2)$ for unambiguous measurements when $\varepsilon = 1$, to the quasi-probability $q(s_1,s_2)$ in the limit $\varepsilon \rightarrow 0 $ of a very weak measurement. Hence the desired quasi-probability is therefore obtained either by a weak measurement, or more generally by using an ambiguous measurement of arbitrary strength together with a projective measurement to determine $p_{12} (s_1, s_2)$. Alternatively, one could use two ambiguous measurements of different strengths $\varepsilon, \varepsilon'$ and use Eq.(\ref{main}) to determine both  $p_{12} (s_1, s_2)$ and $q(s_1,s_2) $ in terms of
$\tilde p_\varepsilon (s_1,s_2) $ and $ \tilde p_{\varepsilon'} (s_1,s_2)$.

These results show quite generally that the quasi-probability Eq.(\ref{quasi}) for {\it any} dichotomic variable $Q$ arises very naturally in a weak or ambiguous measurement scheme, not just in the context of the arrival time problem. Weak measurements in the specific context of the arrival time problem were previously considered in  Ref.\cite{RuKa}, with essential agreement with our results. (See also Ref.\cite{MaRe} for connections between the quasi-probability and weak measurements).

Specific experiments implementing both of the above approaches could be designed using simple modifications of existing protocols designed to measure arrival times. An example is described in Ref.\cite{MBDE}, which involves an incoming packet moving along the $x$-axis interrupting a laser beam in the $z$-direction.


\section{Relation to Standard Arrival Time Distribution Results}

In general the probability $p(t_1,t_2)$ for a free particle crossing the origin during interval $[t_1,t_2]$ has the form
\beq
p(t_1,t_2) = \int_{t_1}^{t_2} dt \ \Pi(t)
\eeq
for some arrival time distribution $\Pi (t)$, for which a number of different candidates have been suggested.
Most approaches yield the quantum-mechanical current in the semiclassical limit,
\bea
\Pi (t) &=& - \frac{i\hbar}{2m}[\psi^*(0,t)\psi'(0,t) - \psi(0,t)\psi'^*(0,t)],
\nonumber \\
&=& \langle \psi | \hat J(t) | \psi \rangle,
\label{cur}
\eea
where the dash denotes spatial derivative and $\hat J$ is the current operator,
\beq
\hat J = \frac{ 1 } {2 m} \left( \hat p \delta ( \hat x) + \delta ( \hat x) \hat p \right).
\eeq
Note for future reference that
\beq
\hat J(t) = \frac{d}{dt} \theta( \hat x(t) ).
\label{diff}
\eeq
For situations in which the wave function vanishes at $x=0$, the expected result is the kinetic energy density,
\bea
\Pi (t) &=& N \langle \psi | \hat p \delta ( \hat x(t) ) \hat p | \psi \rangle 
\nonumber \\
&=& N \hbar^2 \left| \psi'(0,t) \right|^2
\label{strong}
\eea
for some normalization factor $N$, which is typically model-dependent \cite{kin1,kin2,kin3}.
We expect our quasi-probability to make some link to these standard results, at least in the small time limit, and this we now show.

The quasiprobability $q(-,+)$ can be written as
\begin{align} \label{Re}
q(-,+)= {\rm Re} \langle \Psi | {P}_+(\tau) {P}_- | \psi \rangle,
\end{align}
where $\Re$ denotes the real part and we have chosen $t_2 = \tau$ and $t_1 = 0$. Since
\bea
P_+(\tau) P_- &=&(P_+(\tau)-P_+)P_-
\nonumber \\
&=&  \int_0^\tau dt \hat J(t) P_-,
\eea
where we have used Eq.(\ref{diff}), we have
\begin{align}\label{qt}
q(-,+)=\int_0^\tau dt \ \Re\mel{\psi}{\hat{J}(t)P_-}{\psi}.
\end{align}
To evaluate the matrix element we note that
\bea\label{el}
\mel{\psi}{e^{\frac{i}{\hbar}Ht}\hat{J}e^{-\frac{i}{\hbar}Ht}P_-}{\psi}&=&\frac{1}{2m}(\mel{\psi}{e^{\frac{i}{\hbar}Ht}\hat p}{0}\mel{0}{e^{-\frac{i}{\hbar}Ht}P_-}{\psi}
\nonumber \\ &+&\mel{\psi}{e^{\frac{i}{\hbar}Ht}}{0}\mel{0}{\hat p e^{-\frac{i}{\hbar}Ht}P_-}{\psi}).
\eea
We may then use the wave function notations,
\bea
\mel{0}{e^{-\frac{i}{\hbar}Ht}}{\psi}&=&\psi(0,t),
\\
\mel{0}{\hat p e^{-\frac{i}{\hbar}Ht}}{\psi}&=&- i \hbar \frac{ \partial \psi}{ \partial x} (0,t),
\eea
to write this as,
\bea 
\label{maint}
\mel{\psi}{e^{\frac{i}{\hbar}Ht}\hat{J}e^{-\frac{i}{\hbar}Ht}P_-}{\psi}\nonumber
&=&\frac{i \hbar }{2m}\mel{0}{e^{-\frac{i}{\hbar}Ht}P_-}{\psi}\frac{\partial \psi^*(0,t)}{\partial x}
\nonumber \\
&-& \frac{i \hbar }{2m} \psi^*(0,t)\left[ \frac{\partial}{\partial x} \mel{x}{e^{-\frac{i}{\hbar}Ht}P_-}{\psi}\right]_{x=0}.
\eea
The two remaing matrix elements are then evaluated for small $\tau$ by following the methods in Refs.\cite{HELM,Sok}. We thus obtain, at some length, the small-time expansion,
\bea
q(-,+)&=&\frac{1}{2} J(0,0)\tau
\nonumber\\
&+& \frac{(\hbar \tau)^{\frac{3}{2}}}{6\sqrt{\pi}m^{\frac{3}{2}}}\bigg(|\psi'(0,0)|^2-\sqrt{\frac{i}{2}}\psi(0,0)\psi''^*(0,0)-\sqrt{\frac{-i}{2}}\psi^*(0,0)\psi''(0,0)\bigg). \label{finalapprox}
\eea

The final result connects with the standard formulae given above. The leading order behaviour is essentially the current, Eq.(\ref{cur}), apart from the factor of a half, which is due to the fact that the quasi-probability is for a left-right crossing. Hence there is an agreement with semiclassical notions for short times. 
For situations in which the wave function vanishes at $x=0$, we get a result proportional to the kinetic energy density, Eq. (\ref{strong}).

Note that we do not make any contact with the ideal arrival time distribution of Kijowksi \cite{Kij}. This is not surprising since the Kijowski distribution is always non-negative, yet here, we are working with a quasi-probability.

\section{Relationship to Quantum Backflow}

Quantum backflow is the surprising non-classical effect that a free particle in a state $ \psi(x,t)$ consisting entirely of positive momenta can have a negative current \cite{BrMe,Back}. It means, for example, that the probability 
\beq
p_-(t) = \int_{-\infty}^0 dx \ | \psi(x,t)|^2,
\label{pminus}
\eeq
of remaining in the negative $x$-axis can actually increase for periods of time, even though in the long run it decreases, in accordance with classical intuition. The degree of increase can be determined by considering the flux operator
\beq
\int_{t_1}^{t_2} \hat J(t) = P_+ (t_2) - P_+ (t_1),
\label{flux}
\eeq
restricted to the space of states with positive momentum. Following Bracken and Melloy \cite{BrMe}, and subsequent authors, one considers the eigenvalue equation
\beq
\theta (\hat p ) \int_{t_1}^{t_2} \hat J(t) \  | \lambda\rangle = \lambda | \lambda \rangle.
\eeq
It has been shown that the spectrum lies in the range $[-c_{bm}, 1]$, where $c_{bm}$ is the Bracken-Melloy backflow constant and has been determined numerically to take the approximate value $c_{bm} \approx 0.04$. Furthermore, the eigenstates with negative eigenvalue seem to have a discrete spectrum and those with positive eigenvalues are have a continuous spectrum. Backflow is easily shown to be a consequence of a negative Wigner function \cite{Wig}, or equivalently, due to interference effects. To date it has not been measured experimentally, although at least one proposal to do so exists \cite{BackE}.

For our purposes, backflow states seem like natural states to try in looking for situation with $q(-,+)$ is negative. We note that the eigenvalue equation Eq.(\ref{flux}) may be written,
\beq
\frac{1}{2} \theta (\hat p) \left( \hat Q_2 - \hat Q_1 \right) | \lambda \rangle = \lambda | \lambda \rangle,
\eeq
and that from Eq.(\ref{idn}), the $q(-,+)$ quasi-probability may be written
\bea
q(-, +) &=& \frac{1}{8}   ((2 \lambda + 1)^2 - 1 ),
\nonumber \\
&=& \frac{1}{2} \lambda ( \lambda + 1).
\eea
For a maximal backflow state, this means $q(-,+) \approx -0.02$. This is clearly small in comparison to the lower bound on $q(-,+)$ of $- 0.125 $ (see Eq.(\ref{LB})), suggesting that the most negative values of the quasi-probability arise from superpositions of positive and negative momentum states.

These results also suggest that the measurements of the quasi-probability described earlier may also provide a possible experimental test of backflow, if it is possible to prepare non-trivial initial states with non-negative momenta. The results of such an experiment would clearly be interesting. However, such an approach may be unnecessarily complicated in terms of measuring the backflow effect. The quasi-probability requires measurements of the system at two times, whereas the presence of backflow can be determined by measuring the probability Eq.(\ref{pminus}), which would clearly be simpler.

\section{The Quasi-Probability for Gaussian States}

To illustrate some of the properties of the quasi-probability we compute it explicitly for initial states consisting of superpositions of gaussian wavepackets. Our first rather simple example is a superposition of four gaussians,
\beq
| \psi \rangle = \sum_{s, s'} a_{s s'} | \phi_{s s'} \rangle,
\eeq
where $s,s' = \pm 1 $.
The four states $ | \phi_{s s'} \rangle$ are chosen so that they are strongly localized in $x>0$ or $x<0$ at $t_1=0$ and $t_2 = \tau$. The states $|\phi_{\pm \pm} \rangle $ remain in those regions during that time interval and the states $|\phi_{\pm \mp} \rangle$ cleanly cross the origin. The states are therefore approximately orthogonal and we may take the normalization $\sum_{s, s'} |a_{s s'}|^2 =1$.
The quasi-probability is readily evaluated, since the states are approximate eigenstates of the projector product $P_{s_2} (\tau) P_{s_1}$ and we find
\beq
q(s_1, s_2) \approx | a_{s_1 s_2} |^2,
\eeq
the expected semiclassical answer.

To obtain more interesting examples in which the quasi-probability is negative we need gaussians which are significantly chopped by the projections at either time. We take
\begin{align}
|\psi \rangle=\alpha_-\ket{\phi_-}+\alpha_+\ket{\phi_+},
\end{align}
where $\alpha_\pm$ are complex coefficients  and $ | \alpha_+|^2 + | \alpha_-|^2 = 1$. We choose the wave packets to be,
\beq
\phi_{\pm} (x) = \frac  {1} { (2 \pi \sigma^2)^{\frac{1}{4} }} \exp \left( - \frac { (x \mp L)^2 } { 4 \sigma^2 } \mp \ih p_0  x \right).
\eeq
They are therefore peaked at the points $ \pm L$ and we choose $ L \gg \sigma $ so that they are approximate eigenstates of the projectors $P_{\pm}$ at the initial time. (We elaborate on this approximation below). We have chosen the momenta to be equal and opposite, with $p_0 > 0 $, so that the wave packets approach each other. This means that after a time $ \tau = m L / p_0 $, the wave packets meet at the origin and we have,
\beq
\phi_{\pm} (x, \tau) = \frac  {1} { (2 \pi (\Delta x)_\tau^2)^{\frac{1}{4} }} \exp \left( - \frac { x^2 } { 4 \sigma_\tau^2 } \mp \ih p_0 x - \ih E \tau\right),
\eeq
where $\sigma_\tau^2 = \sigma^2 + i \hbar \tau / (2m)$, $E = p_0^2 / 2m $ and
\beq
(\Delta x)_\tau^2 = \sigma^2 + \frac{ \hbar^2 \tau^2 } { 4 m^2 \sigma^2}.
\eeq

The quasiprobability with $t_1 = 0 $ and $t_2 = \tau$ can be written,
\begin{align}
q(\pm,s)&=\Re \mel{\psi}{P_{s}(\tau)P_{\pm}}{\psi}
\\&=\Re \bigg( |\alpha_{\pm}|^2\mel{\phi_{\pm}}{P_{s}(\tau)}{\phi_{\pm}}+\alpha_{\pm}\alpha_{\mp}^*\mel{\phi_{\mp}}{P_{s}(\tau)}{\phi_{\pm}}\bigg),
\label{2g}
\end{align}
where $s$ takes values of $\pm1$. It is easily seen that
\beq
\mel{\phi_{\pm}}{P_{s}(\tau)}{\phi_{\pm}} = \int_0^\infty dx  \ | \phi_{\pm} (x, \tau) |^2 = \frac{1}{2}
\label{easy}
\eeq
The second term is the overlap integral
\begin{align}
\mel{\phi_{\mp}}{P_{s}(\tau)}{\phi_{\pm}}=
\frac  {1} { (2 \pi (\Delta x)_\tau^2)^{\frac{1}{2} }} \int_0^{\infty} dx\, \exp \left( {-\frac{x^2}{2(\Delta x)_\tau^2} 
\mp 2 is  p_0x} \right).
\label{overlap}
\end{align}
These integrals may be evaluated in terms of the imaginary error function $\text{erfi}(y)$ \cite{Bess} and we find
\beq
\mel{\phi_{\mp}}{P_{s}(\tau)}{\phi_{\pm}}= \frac {1}{2} e^{-y^2} ( 1 \pm i \ \text{erfi}(y) )
\eeq
where $ y = p_0 (\Delta x)_\tau  / \hbar $. We thus obtain for the quasi-probability,
\begin{align}\label{2q}
q(\pm,s)= \frac{1} {2} \Re \bigg( |\alpha_{\pm}|^2+  \alpha_{\pm}\alpha_{\mp}^*e^{-y^2}(1\mp i \ \text{erfi}(y))\bigg),
\end{align}
We will suppose that the parameter $y$ is fixed and find values of the coefficients which minimize the quasi-probability.

The complex coefficients can be parameterised as follows
\begin{align}
\alpha_+=\cos\phi\, e^{i\theta_+} && \alpha_-=\sin\phi\, e^{i\theta_-}
\end{align}
with the condition $0\leq\phi\leq\frac{\pi}{2}$. Define $\theta=\theta_+-\theta_-$, then 
\begin{align}
q(+,\pm)&=\frac{1}{2}\cos^2\phi+\frac{1}{4}\sin (2\phi) f_\pm(y,\theta),
\\q(-,\pm)&=\frac{1}{2}\sin^2\phi+\frac{1}{4}\sin (2\phi) f_\pm(y,\theta).
\end{align}
where
\beq 
f_\pm(y,\theta)= e^{-y^2} \left( \cos \theta \pm \sin\theta\, \text{erfi}(y) \right).
\eeq
From the compound angle formula we easily find that for fixed $y$ an angle $\theta$ may be chosen so that $ |f_{\pm}| $ has maximum value,
\beq
f_m (y) = e^{ - y^2} \sqrt{ 1 + {\rm erfi}(y)^2 }.
\eeq
Similarly, the minimum of the quasiprobability can be seen most clearly by expressing it in the form,
\begin{align}\label{qplus}
q(+,\pm)&=\frac{1}{4}[1+\sqrt{1+f^2_\pm}\cos(2\phi-\text{arctan}f_\pm)],
\\q(-,\pm)&=\frac{1}{4}[1-\sqrt{1+f^2_\pm}\cos(2\phi+\text{arctan}f_\pm)].\label{qminus}
\end{align}
which may all be minimized to the value $\frac{1}{4}(1-\sqrt{1+f^2_\pm})$ for some $\phi$. Hence we find that the coefficients in the superposition may be chosen so that the minimum value of the quasi-probability for fixed $y$ is,
\beq
q_{min} (y) = \frac {1}{4} \left( 1 - \sqrt{ 1 + e^{-2y^2} ( 1 + {\rm erfi(y)^2} ) } \right).
\label{qminy}
\eeq
This takes its lowest value of $ \frac{1}{4} ( 1 - \sqrt{2}) \approx -0.104 $ at $y=0$
(which is pretty close to the absolute lower bound on the quasi-probability of $-0.125$) and increases monotonically with $y$.

There is some tension between choosing a small value of $y$ and meeting the requirement $L \gg \sigma$ necessary for the approximation that the states $|\phi_{\pm} \rangle$ are approximate eigenstates of the projectors $P_{\pm}$ and the value $y=0$ is therefore not feasible.
We consider this approximation in more detail. It means that there is an error in the state of the form $P_{\mp} |\phi_{\pm} \rangle$. This means that to leading order there will be an error in the terms in Eq.(\ref{2g}) involving terms of the form $ \mel{\phi_{\pm}}{P_{s}(\tau)P _{\mp}}{\phi_{\pm}}$, which have the upper bound,
\bea
| \mel{\phi_{\pm}}{P_{s}(\tau)P _{\mp}}{\phi_{\pm}}|^2 & \le &
\mel{\phi_{\pm}}{P_{s}(\tau)}{\phi_{\pm}} \mel{\phi_{\pm}} {P_{\mp}} {\phi_{\pm}}
\nonumber \\
&=& \frac{1}{4} \ {\rm erfc} \left( \frac{L} {\sqrt{2} \sigma}\right)
\eea
using Eq.(\ref{easy}), where ${\rm erfc}$ is the complementary error function \cite{Bess}. By plotting (or simply plugging in some numbers), this error is of order $0.005$ (i.e. half a percent error) at $L/\sigma =2.3$, so it does not have to be that large for a small error.

Consider now the possible values of $y$. We have
\beq
y^2 = \frac{p_0^2 \sigma^2}{ \hbar^2} + \frac {L^2} { 4 \sigma^2}
\eeq
where we have inserted the value $\tau = m L/p_0$. It turns out to be convenient to take $\sigma$ to be very small, so now $y \approx L / (2 \sigma)$. With the above value $L/\sigma =2.3$ we have $y=1.15$ and Eq.(\ref{qminy}) then yields the minimum value for the quasi-probability as,
\beq
q_{min} = - 0.05,
\eeq
with an estimated error of $0.005$ due to the approximations, a factor of $10$ smaller.
This minimum value is a reasonable fraction of the most negative quasi-probability of $-0.125$. Hence reasonably negative values for the quasi-probability may be obtained with a simple superposition of gaussians.

A final example consists of a single initial gaussian. The quasi-probability can be written in the momentum space representation as,
\beq
q(s_1,s_2) = \int dp_1 \int dp_2 \ \tilde \psi^* (p_1) \tilde \psi (p_2) f_{s_1 s_2} (p_1, p_2),
\eeq
where 
\beq
f_{s_1 s_2} (p_1, p_2) = {\rm Re} \ \langle p_1 |  P_{s_2} (\tau) P_{s_1}| p_2 \rangle.
\eeq
We concentrate on $q(-,+)$ and take an initial gaussian state very strongly concentrated in momentum about a value $p$ and with zero average position. The quasi-probability is therefore proportional (up to a postive constant) to,
\begin{align}
f_{-+}(p,p) =&\Re\bigg[\sqrt{\frac{m}{2\pi i \hbar \tau}} \int_{-\infty}^0 dx \int_0^\infty dy e^{- \ih p(x-y)} e^{\frac{im(x-y)^2}{2 \hbar \tau}} e^{-i\frac{p^2 \tau}{2m \hbar }} \bigg], \nonumber
\\=&\ \tau \ \Re \bigg[\frac{p}{2m \hbar}\text{erfc}\bigg(-p\sqrt{\frac{i\tau}{2m \hbar }}\bigg)+\frac{1}{\sqrt{i2\pi m \hbar \tau}} e^{-\frac{i p^2 \tau}{2m \hbar }}\bigg],\label{pq}
\end{align} 
where ${\rm erfc}(z)$ is the complementary error function.
As shown in Ref.\cite{RuKa}, this quantity approaches the expected classical value $p\tau/m$ for large positive $p$, but can have a small amount of negativity for small negative $p$. The timescale for the transition from negativity to postivity as $\tau$ increases from zero is determined by the argument of the error function, and is readily seen to be $\tau_E = \hbar / E $, where $E = p^2/ 2m$. This is precisely
the energy time mentioned in the Introduction, at which quantum-mechanical effects in sequential measurements become important \cite{HaYe1}.

\section{The Quasi-probability in the Wigner-Weyl Representation}

Some useful insights into the properties of the quasi-probability may be obtained using the Wigner-Weyl representation, in which hermitian operators are mapped into real phase space functions, according to the transform,
\begin{align}\label{wsym}
W_A(X,p)=\frac{1}{2\pi \hbar}\int_{-\infty}^{\infty} d\xi e^{-ip\xi}\  \mel{X+\frac{\xi}{2}}{\hat{A}}{X-\frac{\xi}{2}}.
\end{align}
The transform of the density operator yields the well-known Wigner function \cite{Wig}. The relevant properties of this transform are conveniently summarized in Ref.\cite{Hal4}. A standard property that we will make use of is the relation,
\beq
{\rm Tr} ( \hat A \rho) = 2 \pi \hbar \int dX \int dp \ W_A (X,p) \ W_{\rho} (X,p).
\eeq
The quasi-probability Eq.(\ref{quasi}) clearly has the form of the left-hand side, so is readily written in Wigner-Weyl form,
\beq
q(s_1, s_2) =    2 \pi \hbar \int dX \int dp \ W_{s_1 s_2} (X,p) \ W_{\rho} (X,p),
\eeq
where $ W_{s_1 s_2} (X,p) $ is the Wigner-Weyl transform of the operators
\beq
A_{s_1 s_2} = \half \left( P_{s_2} (t_2) P_{s_1} (t_1) + P_{s_1} (t_1) P_{s_2} (t_2)  \right).
\eeq

\begin{figure} [h]
\includegraphics[width=\textwidth]{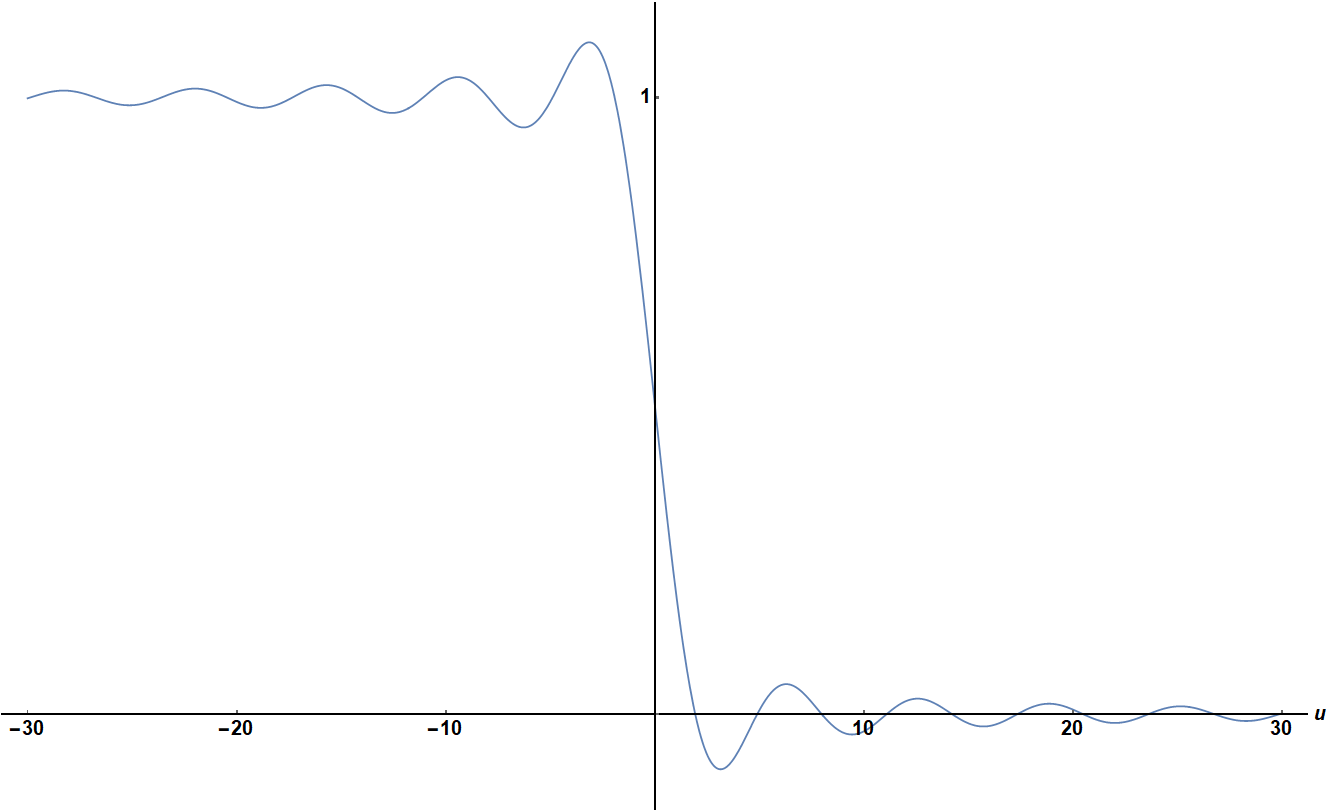}
\centering
\caption{Plot of $\frac{1}{2}-\frac{1}{\pi}\text{Si}(u)$ against $u$, a "quantum step function".}
\label{qstep}
\end{figure}

The transform may be carried out, at some length, with the results,
\begin{align}\label{pmana}
W_{\pm,\pm}(X,p)&=\frac{1}{4\pi}\bigg[\theta(\pm X)+\theta(\pm X_\tau)-\frac{1}{2}+\frac{1}{\pi}\text{Si}\bigg(\frac{2mX}{\tau}X_\tau\bigg)\bigg],
\\W_{\pm,\mp}(X,p)&=\frac{1}{4\pi}\bigg[\theta(\pm X)+\theta(\mp X_\tau)+\frac{1}{2}-\frac{1}{\pi}\text{Si}\bigg(-\frac{2mX}{\tau}X_\tau\bigg)\bigg],
\end{align}
where for convenience we have set $t_1 = 0$, $t_2 = \tau$, and $X_\tau = X + p \tau / m$. Also, $ {\rm Si} (u)$ is the sine integral \cite{Bess}, defined by
\beq
{\rm Si}(u) = \int_0^u \frac { \sin t} {t} dt.
\eeq
The behaviour of $W_{s_1 s_2} (X,p)$ is best seen by plotting   $\frac{1}{2}-\frac{1}{\pi}\text{Si}(u)$, which is shown in
Figure \ref{qstep}. This clearly shows that  $\frac{1}{2}-\frac{1}{\pi}\text{Si}(u)$ is a quantum version of the step function $\theta (-u)$, tending exactly to this $\theta$-function for large $|u|$, but displaying oscillations around $u=0$.
This means that in the regime of large $|u|$, $W_{s_1 s_2} (X,p)$ have the approximate form
\begin{align}\label{pmana2}
W_{\pm,\pm}(X,p)&=\frac{1}{4\pi}\bigg[\theta(\pm X)+\theta(\pm X_\tau)- \theta(- X X_\tau) \bigg],
\\W_{\pm,\mp}(X,p)&=\frac{1}{4\pi}\bigg[\theta(\pm X)+\theta(\mp X_\tau)+\theta (X X_\tau) \bigg].
\end{align}
This is the expected form in the classical regime. 

These results show that $W_{s_1 s_2}(X,p)$ is non-negative throughout most of the phase space space. The only way it can be negative is in the small $u$ regime, as long that the $\theta( \pm X$ and $\theta (\pm X_\tau)$ terms make negligible contribution.
There are then two ways in which the quasi-probability can end up being negative. The first and most obvious way is for the Wigner function $W_{\rho}(X,p)$ to be negative somewhere. This will be the case for any non-gaussian state and indeed we have seen that superpositions of gaussians (which have negative Wigner function) can produce a negative quasi-probability. The other way is to choose a gaussian initial state, so the Wigner function is positive, but arrange for it to be concentrated around the region where one of the quantities $W_{s_1 s_2} (X,p)$ is negative, i.e. for small $u$ together with suitable restrictions on $X$ and $X_\tau$. Again we have seen in Section 6 that a single gaussian can achieve negative quasi-probability.

\section{Interpretation of the Quasi-probability when non-negative}

We now address the question as to whether the quasi-probability may be regarded as a genuine probability when non-negative.
Although it clearly satisfies the mathematical requirements, it is measured in a very indirect way and there is no clear association with the relative frequencies of measured results, the most commonly-employed notion of probability.  We argue here that there is such a connection.

As indicated earlier, the quasi-probability becomes of interest when the sequential measurement probability $p_{12}(s_1,s_2)$ (which does have a clear connection to relative frequencies) fails to satisfy the sum rule, Eq.(\ref{cond}), but $q(s_1,s_2) \ge 0 $.
It is then reasonable to ask if their exists a classical model of this situation in which $q(s_1,s_2)$ is the underlying probability, which one attempts to measure using sequential measurements, but
there is an unknown classical disturbance causing the violation of the sum rule.
From the quantum-mechanical perspective, the failure of the sum rule of due to inteference effects. We are therefore asking whether these interference effect can be modelled as a classical disturbance if they are not too large.

To be more precise, we ask if 
the sequential measurement probability can be expressed in the form,
\beq
p_{12}(s_1,s_2') = \sum_{s_2} k_{s_2' s_2} (s_1) \ q (s_1,s_2).
\eeq
Here, $k_{s_2' s_2}(s_1)$ is the probability, for fixed $s_1$, that the value $s_2$ of $Q$ at time $t_2$ is disturbed by the measurement to the value $s_2'$, and satisfies,
\beq
\sum_{s_2'}k_{s_2' s_2} (s_1) = 1.
\label{kprob}
\eeq
The disturbance probability $k_{s_2' s_2}( s_1)$ can be found using the moment expansions Eqs.(\ref{mom}), (\ref{momp}) for $ q (s_1,s_2)$ and $p_{12}(s_1,s_2)$. It will not be unique and it is also not obviously non-negative, so it would be necessary to show that parameters can be chosen to ensure this.
However, there is a more elegant construction, which is to seek a joint probability $p(s_1,s_2,s_2')$ matching $ p_{12}(s_1,s_2')$
and $ q (s_1,s_2) $. This can be carried out using standard methods and the proof of this is given in the Appendix. Its existence then implies that the disturbance probability is
\beq
 k_{s_2' s_2} (s_1) = \frac{ p(s_1,s_2,s_2')}{ q(s_1,s_2)},
\label{job}
\eeq
which is readily seen to do the job.

Hence we see that the quasi-probability does have a link to relative frequences -- it is the underlying probability distribution in a measurement process consisting of sequential measurements with an unavoidable disturbance.

\section{The Arrival Time Problem as  a Leggett-Garg Test of Macrorealism}

So far we have envisaged experiments which simply measure the quasi-probability to determine whether it is positive or negative in certain situations, thereby checking the predictions of quantum mechanics.
 However, one can design stricter experiments which go much further and, in addition, rule out all plausible classical explanations of the data.
This is precisely what the Leggett-Garg (LG) framework was designed to do.

The LG framework tests a very specific and rather weak notion of classicality called {\it macrorealism}, which loosely speaking, is the notion that the system can be asserted to possesses definite properties at a sequence of times, independent of past or future measurements \cite{LG1}. (See Ref.\cite{ELN} for a useful review and Ref.\cite{MaTi} for a critique).
This notion is made more precise by breaking it into three separate assumptions: (1) the system is in one of the states available to it at each moment of time (macrorealism per se); (2) it is possible in principle to determine the state of the system without disturbing the subsequent dynamics (non-invasive measurability, NIM); (3) future measurements cannot affect the present state. When applied to our dichotomic variable $Q$ measured at a sequence of times, these assumptions guarantee the existence of an underlying joint probability distribution for the results of the measurements, which in turn implies the existence of a set of inequalities similar in form to the Bell and CHSH inequalities.

Here, we consider measurements at just two moments of time and denote by $Q_1$ and $Q_2$ the values of $Q$ at times $t_1$ and $t_2$. In a macrorealistic theory, $Q_1, Q_2$ take definite values $\pm 1 $ which means that,
\beq
(1 + s_1 Q_1) (1 + s_2 Q_2 ) \ge  0,
\eeq
where $s_1, s_2 = \pm 1 $.
The existence of a joint probability means we can simply average these four equations and we obtain the four two-time LG inequalities,
\bea
1 +  \langle Q_1 \rangle + \langle Q_2 \rangle + C_{12} &\ge& 0, 
\label{D1}\\
1 -  \langle Q_1 \rangle - \langle Q_2 \rangle + C_{12} &\ge& 0, \\
1 +  \langle Q_1 \rangle - \langle Q_2 \rangle - C_{12} &\ge& 0, \\
1 -  \langle Q_1 \rangle + \langle Q_2 \rangle - C_{12} &\ge& 0,
\label{D4}
\eea
where $C_{12} = \langle Q_1 Q_2 \rangle $. These are precisely the four conditions of non-negative quasi-probability $q(s_1, s_2) \ge  0 $.
More general (three and four-time) LG inequalities are readily obtained when measurements at three or more times are contemplated.

What is new in the LG framework compared to the measurements contemplated in earlier sections
is the way the quantities are measured -- they must respect the NIM requirement, otherwise the disturbances produced by the measurements could be used to construct classically invasive models which replicate the quantum results \cite{LGinv}. This is trivial for
$\langle Q_1 \rangle $ and $\langle Q_2 \rangle$ which are simply measured in two different experiments. It is non-trival for $C_{12}$ which is measured in a third experiment using a pair of sequential measurements. In this case,
NIM is normally implemented using {\it ideal negative measurements} at the first time in which the detector is coupled to, say, the $Q=-1$ state only, and a null result is taken to mean that the system is in the $Q=+1$ state.  From a macrorealistic point of view this eliminates alternative classical explanations.
Experiments implementing this requirement have been successfully carried out \cite{Knee,Rob,Kat}.

Hence the arrival time problem formulated as a Leggett-Garg test will not only be able to confirm the predictions of quantum mechanics. It will also rule out alternative classical explanations and is therefore a more decisive test of quantumness.


Recall that in Section 3(B) we also used ambiguous measurements to determine the quasi-probability, which includes weak measurements as a special case. It may seem that such measurements are 
non-invasive in the weak limit, since the disturbance to the system becomes arbitarily small -- the inferred probability Eq.(\ref{main}) will satisfy the probability sum rule Eq.(\ref{cond}) arbitrarily well as $\varepsilon \rightarrow 0 $. However, this is perhaps misleading since Eq.(\ref{main}) has been obtained through an inversion procedure involving the inverse matrix $d_{s \alpha}$ which, as one can see from Eq.(\ref{mat}) is singular in the weak measurement limit. It is perhaps clearer to consider instead the ambiguously determined probability Eq.(\ref{ambp}), which will also satisfy the sum rule arbitarily well for small $\varepsilon$, but from expanding out this probability, one can see that the desired quasi-probability is read off from a term which is the same order of magnitude as the sum rule violation. This is a common feature of weak measurements, although some variants have been proposed in which it is claimed that the magnitude of the disturbance is significantly smaller than the effect sought after.
Despite the fact that numerous experimental LG tests have been based on weak measurements (e.g. the very first experimental test of the LG inequalities \cite{Pal}) question marks still remain around whether they really satisfy the NIM requirement. (See for example, the closing remarks in the review article Ref.\cite{ELN}). We will therefore not assume this here.


The lower bound of $- \frac{1}{8}$, Eq.(\ref{LB}), on the quasi-probability $q(s_1,s_2)$ corresponds to a lower bound of $- \frac{1}{2}$ on the right-hand side of the LG inequalities above, which is the familiar Tsirelson bound \cite{Tsi}. Although note that this can in fact be violated in some circumstances in LG experiments by ``degeneracy breaking" measurements \cite{BuEm}.

All LG tests carried out to date concern systems with discrete variables, usually simple spin systems. What is new here in comparison is that the dichotomic variable $Q$ is the sign function of a continuous variable. LG tests on such systems are yet to be explored.

Finally, note that examples of quantum states violating the LG inequalities are supplied by the examples of states with negative quasi-probability discussed in Sections 5 and 6. In particular, it means that states with backflow can violate the LG inequalities, hence there is a coincidence between two quite different notions of non-classical behaviour.

\section{Summary and Discussion}

The purpose of this paper was to explore the quasi-probability Eq.(\ref{quasi}) as both a candidate  measurement-independent probability for the arrival time problem and indicator of quantum-mechanical behaviour. We explored its general properties in Section 2 and demonstrated its relationship to the usual sequential measurement formula. The quasi-probability is non-negative as long as quantum interference is suitably bounded.  When negative it has a lower bound of $- \frac{1}{8}$.

We showed in Section 3 how it can be measured, either indirectly by measuring its moments, or directly, using weak or ambiguous measurements. We saw, in particular, that the quasi-probability Eq.(\ref{quasi}) for any system with a dichotomic variable $Q$  has a natural link to weak measurements (a fact previously noted in the context of the arrival time problem in Ref.\cite{RuKa}).

In Section 4 we showed how the quasi-probability coincides with standard results involving the current and kinetic energy density in the short time limit.
Specific types of states exhibiting negative quasi-probability were exhibited in Sections 5 and 6, namely states with backflow, superpositions of gaussians, and a single gaussian. Some insight into the general conditions under which the quasi-probabilty could be negative was given in Section 7, using the Wigner-Weyl representation.

Our study of the quasi-probability was partly motivated by the possibility that, when non-negative, it may provide a probabilty for the arrival time problem. The extent to which this may be the case was explored in Section 8. We showed that, although it does not have a direct connection with the relative frequencies of certain measured result, it can, when non-negative, be interpreted as the underlying probability in a classical disturbing measurement process.

In Section 9, we saw that the four inequalities for the quasi-probabiltiy $ q(s_1, s_2) \ge 0 $ have the precise form of a set of two-time Leggett-Garg inequalities and the connection with the LG framework was described. This framework goes beyond merely detecting the presence of quantum coherence -- it checks for  whether macrorealistic descriptions can be ruled out.  Simple modifications of the experimental measurement of the quasi-probability were described which meet the requirements of an LG test.

We have focused in this paper on the case of arrival times defined by just two position measurements. Although the simplest possible case, it nevertheless gives interesting results in terms of exploring the presence or absence of quantumness in the arrival time problem. The natural generalization to consider is
$n$-time quasi-probability (given in Ref.\cite{GoPa}),
\beq
q(s_1, s_2, \cdots s_n) = {\rm Re} {\rm Tr} \left(  P_{s_n} (t_n) P_{s_{n-1}} (t_{n-1}) \cdots  P_{s_1} (t_1) \rho\right),
\eeq
a clear analogue of the $n$-time measurement formula Eq.(\ref{1.2}). We would expect this formula also to have a natural connection to sequences of weak measurements.
Interestingly, the quasi-probability approach and the LG inequalities no longer coincide for three or more times -- the requirement of non-negative $n$-time quasi-probability is in fact stronger than the corresponding LG inequalities. This was explored in Ref.\cite{HaYe2}. It means that there are ways of determining an underlying probability that are weaker than using an explicit quasi-probability expression. These possibilities will be explored in future publications.

\section{Acknowledgements}

We are grateful to Gonzalo Muga and James Yearsley for many useful conversations on these topics over a long period of time.

\appendix

\section{Joint Probability for Disturbing and Non-Disturbing Measurements}

In this appendix we show how to construct the joint probability distribution used in Eq.(\ref{job}). The probability $q(s_1,s_2)$ (here taken to be non-negative) is the probability for the variables $Q$ at times $t_1$ and $t_2$, denoted $Q_1$ and $Q_2$, where it is assumed that $Q_2$ is measured in such a way that there is no disturbance from the earlier measurement, so represents that actual value of $Q$ at time $t_2$.
It has moment expansion
\beq
q(s_1,s_2) = \frac {1}{4} \left(1 + \langle Q_1 \rangle  s_1 +  \langle  Q_2 \rangle s_2  + C_{12} s_1 s_2 \right).
\eeq
The probability $p_{12}(s_1,s_2)$ however, represents the situation in which the value of $Q$ at time $t_2$ is disturbed by the earlier measurement. A convenient way to describe the situation, following the contextuality by default approach of Ref.\cite{DzKu}, is to declare that $Q$ at time $t_2$ in the latter context should be regarded as a different variable denoted $Q_2^{(1)}$, since it is measured in a different context,
but is probabilistically related to $Q_2$. We thus write the moment of expansion as
\beq
p_{12} (s_1,s_2') = \frac {1}{4} \left(1 + \langle Q_1 \rangle  s_1 +  \langle  Q_2^{(1)} \rangle s_2'  + C'_{12} s_1 s_2' \right).
\eeq
Here we are also allow the possibility that the correlation functions of the above two probabilities are different but we are ultimately interested in the case in which they are equal. We also introduce the joint probability for $Q_2$ and $Q_2^{(1)}$,
\beq
p_m (s_2,s_2') = \frac {1}{4} \left(1 + \langle Q_2 \rangle  s_1 +  \langle  Q_2^{(1)} \rangle s_2'  + C_{22} s_2 s_2' \right),
\label{Ccon1}
\eeq
for some correlation function $C_{22}$ which is chosen to make $p_m  (s_2, s_2')$ non-negative, so must satisfy
\beq
-1 + | \langle Q_2 \rangle + \langle Q_2^{(1)} \rangle | \le C_{22} \le 
1 - | \langle Q_2 \rangle - \langle Q_2^{(1)} \rangle |.
\eeq
These three pairwise probabilities are all compatible with each other, so satisfy conditions of the form,
\beq
\sum_{s_2} q(s_1, s_2) = \sum_{s_2'} p_{12} (s_1,s_2'),
\label{comp}
\eeq
plus two more similar sets of conditions.

The question now is whether we can find a joint probability distribution $p(s_1,s_2,s_2')$ on all three variables $Q_1, Q_2, Q_2^{(1)}$ which matches the above three pairwise marginal distributions.
A candidate distribution is readily written down using a moment expansion,
\bea
p(s_1,s_2,s_2') &=& \frac{1}{8} \left( 1 + \langle Q_1 \rangle s_1 +  \langle Q_2 \rangle s_2 + \langle Q_2^{(1)} \rangle s_2' 
\right.
\nonumber \\
&+& \left.   C_{12} s_1 s_2  +  C_{12}'s_1 s_2'   +  C_{22}s_2 s_2'  +  {D}s_1 s_2 s_3 \right)
\label{pm}
\eea
Here $D$ is the triple correlator and is not fixed by the pairwise probabilities. This object clearly matches the three marginals but is not necessarily non-negative. The question is then whether a value of $D$ can be chosen to ensure non-negativity of $p (s_1,s_2,s_2') $. The answer to this question is supplied by Fine's theorem \cite{Fine} (stated in a more notationally convenient form in Ref.\cite{HalF}). This theorem ensures that a value of $D$ can be found ensuring non-negativity as long as the three marginals are non-negative (which we assume), the compatibility conditions of the form Eq.(\ref{comp}) hold (which is clearly true) and most importantly, as long as
the following four Bell/LG inequalities hold:
\bea
1 + C_{12} + C_{12}' + C_{22} &\ge& 0,
\\
1 + C_{12} - C_{12}' - C_{22} &\ge& 0, 
\\
1 -  C_{12} + C_{12}' - C_{22} &\ge& 0, 
\\
1 - C_{12} - C_{12}' + C_{22} &\ge& 0. 
\eea
The case we are interested in is that in which $C_{12} = C_{12}'$, in which case the Bell/LG inequalities reduce to the condition
\beq
C_{22} \ge  -1 +  2 | C_{12} |.
\label{Ccon2}
\eeq
We now need to show that $C_{22}$ can be chosen to satisfy the two conditions Eq.(\ref{Ccon1}) and Eq.(\ref{Ccon2}). This requires the lower bound in Eq.(\ref{Ccon2}) to be consistent with the upper bound in Eq.(\ref{Ccon1}), which is equivalent to requiring
\beq
2 - | \langle Q_2 \rangle - \langle Q_2^{(1)} \rangle |  - 2 | C_{12} | \ge 0.
\eeq
This does in fact hold using the fact that
\beq
q(s_1, s_2) + p_{12} (-s_2,-s_2) \ge 0.
\eeq
Hence a suitable value of $C_{22}$ can be chosen. This completes the proof of the existence of joint probability distribution.

\end{document}